\makeatletter
\let\@acmBadgeL@image\@empty
\let\@acmBadgeR@image\@empty
\makeatother
\documentclass[sigconf]{acmart}
\AtBeginDocument{%
 \providecommand\BibTeX{{%
   \normalfont B\kern-0.5em{\scshape i\kern-0.25em b}\kern-0.8em\TeX}}}

\usepackage{colortbl}
\usepackage{xspace}
\newcommand{\ie}{\textit{i.e., \xspace}}
\newcommand{\eg}{\textit{e.g., \xspace}}
\newcommand{\etal}{\textit{et al. \xspace}}
\usepackage{array}
\newcolumntype{L}{>{\arraybackslash}m{16cm}}
\newcolumntype{C}[1]{>{\centering\let\newline\\arraybackslash\hspace{0pt}}m{#1}}
\newcolumntype{R}[1]{>{\raggedleft\let\newline\\arraybackslash\hspace{0pt}}m{#1}}
\usepackage[numbered]{bookmark}
\usepackage{tcolorbox}
\usepackage{graphicx}
\usepackage[switch]{lineno}
\usepackage{lineno}
\usepackage{xcolor}
\usepackage{stfloats}
\usepackage[justification=centering]{caption}
\usepackage{dirtytalk}
\usepackage{amsmath}
\usepackage{pgfplots, pgfplotstable}
\usetikzlibrary{patterns}

\usepackage[english]{babel}
\usepackage{textcomp}
\usepackage{mathtools}
\usepackage{ltxtable}
\usepackage{algorithm}
\usepackage{algpseudocode}
\usepackage{textcomp}
\usepackage{tcolorbox}

\usepackage{pgf-pie}
\definecolor{wedge1}{RGB}{ 190  30  46}
\definecolor{wedge2}{RGB}{ 240  65  54}
\definecolor{wedge3}{RGB}{ 241  90  43}
\definecolor{wedge4}{RGB}{ 247 148  30}
\definecolor{wedge5}{RGB}{  43  56 144}
\definecolor{wedge6}{RGB}{  28 117 188}
\definecolor{wedge7}{RGB}{  40 170 225}
\definecolor{wedge8}{RGB}{ 119 179 225}
\definecolor{wedge9}{RGB}{ 181 212 239}
\definecolor{wedge10}{RGB}{  0 104  56}
\definecolor{wedge11}{RGB}{  0 148  69}
\definecolor{wedge12}{RGB}{ 57 181  74}
\definecolor{wedge13}{RGB}{141 199  63}
\definecolor{wedge14}{RGB}{215 244  34}
\definecolor{wedge15}{RGB}{249 237  50}
\definecolor{wedge16}{RGB}{248 241 148}
\definecolor{wedge17}{RGB}{242 245 205}
\definecolor{wedge18}{RGB}{123  82  49}
\definecolor{wedge19}{RGB}{104  73 158}
\definecolor{wedge20}{RGB}{102  45 145}
\definecolor{wedge21}{RGB}{148 149 151}
\definecolor{wedge22}{RGB}{ 204 50 153}
\definecolor{wedge23}{RGB}{ 79 47 79}
\definecolor{wedge24}{RGB}{ 173 234 234}
\definecolor{wedge25}{RGB}{ 216 191 216}
\definecolor{wedge26}{RGB}{  43  56 144}
\definecolor{wedge27}{RGB}{  40 170 225}
\definecolor{wedge28}{RGB}{ 119 179 225}
\definecolor{wedge29}{RGB}{ 181 212 239}
\definecolor{wedge30}{RGB}{  0 104  56}
\definecolor{wedge31}{RGB}{  0 148  69}
\definecolor{wedge32}{RGB}{ 57 181  74}

\usetikzlibrary{chains}
\pgfmathsetmacro\startAngle{90-3.6/2}
\pgfmathsetmacro\radius{+5}
\pgfmathsetmacro\maxLeg{+12}
\pgfmathsetmacro\legBound{+60}
\pgfmathsetmacro\legSpacing{2*\legBound/(\maxLeg-1)}
\usepackage{verbatim}
\pgfplotsset{compat=1.14}
\usetikzlibrary{patterns,}

\usepackage{graphicx}
\usepackage{tabularx,booktabs}
\usepackage{array}
\usepackage{bbding}
\usepackage{pifont}
\usepackage{wasysym}
\usepackage{caption}
\usepackage{dirtytalk}
\usepackage{subcaption}
\usepackage{tabularx}
\usepackage{sfmath}
\usepackage{array, booktabs, makecell}
\usepackage{color}
\usepackage{csquotes}
\usepackage{url}
\usepackage{comment}
\usepackage{tikz}
\usepackage{balance}
\usepackage{listings}
\usepackage{booktabs} % used for \toprule in tables
\usepackage{soul} % highlighting
\usetikzlibrary{fit} %% Used for putting dotted box in image
\usetikzlibrary{positioning}
\usetikzlibrary{arrows}
\usetikzlibrary{shapes.multipart}
\usepackage{adjustbox}
\usepackage{float}
\usepackage{rotating}
\usepackage{nth}
\DeclareCaptionType{TextBox}
\usepackage{pstricks}
\usepackage{placeins}
\usepackage{afterpage}
\usepackage{multirow} 
\usepackage{textcomp}
\usepackage{multicol}
\usepackage{varwidth} % Double column tables
\newcommand{\toolname}{\textsc{Refactoria}\xspace}
\author{Eman Abdullah AlOmar}
\orcid{0000-0003-1800-9268}
\affiliation{%
  \institution{Stevens Institute of Technology}
  \city{Hoboken}
  \state{New Jersey}
  \country{USA}
}
\email{ealomar@stevens.edu}

\author{Christopher Engelbart}
\orcid{0009-0008-1498-756X}
\affiliation{%
  \institution{Stevens Institute of Technology}
  \city{Hoboken}
  \state{New Jersey}
  \country{USA}
  }
\email{cengelba@stevens.edu}

\author{Stephen Pachucki}
\orcid{0009-0008-0257-3394}
\affiliation{%
   \institution{Stevens Institute of Technology}
  \city{Hoboken}
  \state{New Jersey}
  \country{USA}
}
\email{spachuck@stevens.edu}

\author{Nerissa Lundquist}
\orcid{0009-0001-2890-8286}
\affiliation{%
  \institution{Stevens Institute of Technology}
  \city{Hoboken}
  \state{New Jersey}
  \country{USA}
 }
\email{nlundqui@stevens.edu}

\author{David Van Hise}
\orcid{0009-0008-1413-118X}
\affiliation{%
   \institution{Stevens Institute of Technology}
  \city{Hoboken}
  \state{New Jersey}
  \country{USA}
  }
\email{dvanhise@stevens.edu}

\begin{document}

%\title{Reflections on teaching code refactoring}
\title{Leveraging Game-Based Platform to Teach Code Refactoring: An Experience with \toolname}


%\renewcommand{\shortauthors}{Anon, et al.}

\begin{abstract}
Refactoring is the art of improving the internal structure of the code without altering its external behavior. Because of the topic's significance, several teaching methods and strategies have been proposed in the literature. However, skills in identifying and refactoring code smells come from training and experience, and a lack of motivation may hinder developers' adoption of refactoring tools.  In this paper, we discuss the results of an experiment in the classroom that involved performing various refactoring activities to remove antipatterns using \toolname, an innovative game-based tool that supports the acquisition of code smell and refactoring concepts. The players play as an expert chef with their sidekick Watson the Whiskbot to refactor Watson's instructions into efficient, readable, and easily maintainable code. We present an experiment with 30 student developers. In particular, we study the perception, effectiveness, and usefulness of gamification for engaging developers in code smell identification and refactoring.  Our evaluation indicates a high perceived usefulness among students, and participants reported that \toolname facilitated their understanding of the refactoring concepts. %Our results show that the game is effective in teaching developers code smells and refactoring concepts. 

\end{abstract}


\begin{comment}

\begin{CCSXML}
<ccs2012>
 <concept>
  <concept_id>10010520.10010553.10010562</concept_id>
  <concept_desc>Computer systems organization~Embedded systems</concept_desc>
  <concept_significance>500</concept_significance>
 </concept>
 <concept>
  <concept_id>10010520.10010575.10010755</concept_id>
  <concept_desc>Computer systems organization~Redundancy</concept_desc>
  <concept_significance>300</concept_significance>
 </concept>
 <concept>
  <concept_id>10010520.10010553.10010554</concept_id>
  <concept_desc>Computer systems organization~Robotics</concept_desc>
  <concept_significance>100</concept_significance>
 </concept>
 <concept>
  <concept_id>10003033.10003083.10003095</concept_id>
  <concept_desc>Networks~Network reliability</concept_desc>
  <concept_significance>100</concept_significance>
 </concept>
</ccs2012>
\end{CCSXML}
\end{comment}

\begin{CCSXML}
<ccs2012>
   <concept>
       <concept_id>10011007.10011006.10011073</concept_id>
       <concept_desc>Software and its engineering~Software maintenance tools</concept_desc>
       <concept_significance>500</concept_significance>
       </concept>
   <concept>
       <concept_id>10011007.10011074.10011111.10011696</concept_id>
       <concept_desc>Software and its engineering~Maintaining software</concept_desc>
       <concept_significance>500</concept_significance>
       </concept>
 </ccs2012>
\end{CCSXML}

\ccsdesc[500]{Software and its engineering~Software maintenance tools}
\ccsdesc[500]{Software and its engineering~Maintaining software}

\keywords{refactoring, code smell, quality, software engineering, education}
\maketitle
\section{Introduction}
\label{Section:Introduction}

Refactoring is the art of remodeling the software design 
without changing its external behavior \cite{Fowler:1999:RID:311424,mens2004survey,alomar2021preserving}. When performed effectively, it leads to significant software quality improvement. Thus, refactoring is recommended practice according to the software engineering body of knowledge \cite{bourque2002guide}.  Being critical to software quality, refactoring research has been of importance to practitioners, researchers, and educators. Despite the fact that refactoring is considered useful and important,  surveys have shown a lack
of adoption of refactoring tools in practice \cite{Silva2016why,murphy2012we,kim2012field}, \textcolor{black}{highlighting} the importance of teaching software engineering tools to students \cite{raibulet2019teaching}. As large language models (LLMs) are increasingly used for code generation, maintaining code quality is crucial, as AI-generated code often exhibits quality issues \cite{carneiro2025can,yeticstiren2023evaluating}.

Refactoring is a critical software maintenance activity that developers perform for a number of reasons \cite{Silva2016why,alomar2021ESWA,kim2012field,Tsantalis:2013:MES:2555523.2555539,murphy2012we,palomba2017exploratory,szHoke2017empirical}. One of the main refactoring motivations is code smell resolution. Code smell is any characteristic of software that might indicate a problem, negatively affecting software quality \cite{Fowler:1999:RID:311424}. The skills needed to identify and refactor code smells come from training and experience \cite{haendler2019serious,dos2019cleangame}, and lack of motivation can
hinder developers in the adoption of refactoring tools \cite{elezi2016game}. Although these advantages exist, skills in code smells and refactoring have been neglected in some computer science or software engineering courses \cite{dos2019cleangame}, as these topics are generally less popular, compared to design patterns.  In many computer science or software engineering curricula, instructors highlight software quality when teaching good programming practices and design patterns. However, it is necessary to reveal how deviation from best practices can lead to poor design choices that negatively impact the source code \cite{alomar2024automating}. Due to the importance of the topic, several methods and strategies for teaching and training refactoring have been proposed including interactive learning \cite{keuning2021tutoring,haendler2019interactive,keuning2020student,haendler2019refactutor,raab2012codesmellexplorer,alomar2024automating}, incremental learning \cite{smith2006innovative,stoecklin2007teaching}, gamification \cite{elezi2016game,haendler2019serious,dos2019cleangame,aljedaani2024boring}, visualization \cite{raab2012codesmellexplorer}, experimentation \cite{abid2015reflections,izu2022resource,demeyer2005lan,tsantalis2018ten,tsantalis2009identification,tsantalis2015assessing,menolli2024teaching,alomar2024automating,bezerra2026refactoring}, and e-activity planning \cite{lopez2014design}. Although these methods contribute to enhancing a student's grasp of refactoring, it is essential to introduce students to more profound design-level antipatterns, which often appear in even the most well-engineered projects \cite{bavota2015experimental} and are more challenging to resolve \cite{palomba2018diffuseness}. A working knowledge of refactoring is imperative to be a proficient programmer, especially when working with large codebases; it is often impossible to efficiently understand and update code if it has not been refactored. However, some developers are unaware of this skill. Furthermore, self-taught developers are frequently also not exposed to refactoring. This results in less-quality code, creating more work for fewer results in the long run.

To cope with the above-mentioned challenges, we propose the adoption
of gamification in a refactoring environment (\ie the application of game elements to a non-game environment \cite{hamari2014does,huotari2012defining}). To encourage learning in this essential skill, we developed \toolname, a novel game-based learning environment that teaches the basic concepts of refactoring. Therefore, the learning goals of \toolname are related to growing as a programmer by understanding indicators of poor code design, and how to fix the underlying issues that those indicators represent. The overall gameplay revolves around editing code; more specifically, the players are given flawed code and told to improve it. This gameplay loop should be easily recontextualized in the real world, as players reapply the lessons they learn in external situations. The narrative, puzzle structure, and learning feedback all contribute to this core learning concept.

The players learn to identify various code smells and apply refactoring techniques to eliminate them. \toolname follows a chef and their cooking robot, Watson the Whiskbot, in their quest to improve Watson’s code base. On each of the thirteen levels, the player is given some code that, when run, accepts a list of orders and directs Watson to cook and deliver each order. Although the code passes the test cases, it is riddled with code smells. The player is then tasked to refactor the code to eliminate the code smells using refactoring techniques while still passing all test cases. At the end of each level, the player is presented with the names of each refactoring technique they used and each code smell that they discovered. Each technique and smell are added to the game dictionary after discovery so that players can reference previously discovered techniques and smells. Techniques and smells build on each other throughout the game.  The material is available online\footnote{\url{https://refactorings.github.io/education/}}.
%https://shorturl.at/Vp1iV

%The remainder of this paper is organized as follows. Section \ref{Section:Background} reviews existing studies related to software refactoring in education. Section \ref{Section:GameDesign} outlines our game design in terms of narrative and design. Section \ref{Section:RefactorLang}
%discusses \toolname programming language, while its puzzles are discussed in Section \ref{Section:puzzles}. Section \ref{Section:Result} captures the game evaluation. The limitations
%of our game are discussed in Section \ref{Section:Threats} before concluding with Section \ref{Section:Conclusion}.

\section{Background and Related Work}
\label{Section:Background}

Demeyer \etal \cite{demeyer2005lan}  proposed the LAN-simulation as an example of a
 refactoring scenario that mimics realistic circumstances. The authors illustrated the applicability of the example by applying various tools supporting different aspects of refactoring. Santos \etal \cite{dos2019cleangame} implemented \textsc{CleanGame}, a gamified tool that helps to identify five code smells: \say{Large Class}, \say{Long Method}, \say{Divergent Changes}, \say{Feature Envy}, and \say{Shotgun Surgery}. The game consists of two modules: (1) smell-related quiz (\ie shows questions about code smell resolution with multiple-choice answers), and (2) code smell identification (\ie presents questions about identifying code smells in the given source code). Their findings reveal that participants managed to identify twice as many code smells with a gamified approach compared
to a non-gamified approach. Haendler \cite{haendler2019card} proposed a non-digital multi-player card game called \textsc{Refactory} to learn the principles of refactoring without development. Their user study indicate that the card game can
encourage learning about refactoring and
support learning refactoring principles (\ie combining refactoring operations to remove code smells). Elezi \etal  \cite{elezi2016game} created \textsc{CodeArena}, a gamification system
that tracks and rewards refactorings during development. While useful, the authors indicated that the game does not perform any quality evaluation on the types refactorings completed, and
 does not evaluate if the refactored code still works by passing test cases. Their results show that gamification had less effect than expected, but can be useful to practitioners interested in promoting refactoring tools via gamification. Baars and Meester \cite{baars2019codearena} proposed \textsc{CodeArena}, an extension to the popular 3D sandbox game called \textsc{Minecraft} \cite{gupta2015minecraft}, that converts patterns in a codebase that are considered harmful for the purpose of improving code quality. When the Java project has been selected, the tool creates a
monster for each code issue found. The user then needs to fix
code issues to collect points. Haendler and Neumann \cite{haendler2019serious} investigated the possibilities and challenges of designing serious games for refactoring on real-world code artifacts by proposing
a game design, where students can compete either against a predefined benchmark (technical debt) or against each other. Agrahari and Chimalakonda \cite{agrahari2020refactor4green} developed \textsc{Refactor4Green}, a desktop game created using the Unity 3D game engine to teach code smells and refactoring
to novice programmers. The central concept of the game involves incorporating code smells via refactoring decisions, all within the theme of an environmentally green setting. The authors focused on code smells that result in more energy consumption of software. Their evaluation show positive feedback from 83.06\% of the participants. Aljedaani \etal \cite{aljedaani2024boring} proposed \textsc{RefGame} to enhance the learning experience by identifying and applying refactoring techniques. The game consists of three different types: matching, racing, and snake games, each highlights on different aspects of refactoring. Their exploratory study with 322 students show  that the proposed game educational tool made software refactoring education accessible
and effective. %Table \ref{tab:related-work-combined} summarizes the code smells and refactoring in game-based refactoring techniques.

%The above-mentioned game-based studies are not publicly available. \textsc{CodeArena} is an exception, but it focuses exclusively on code clones. 
Our work differs from these studies because it engages students in refactoring a full software system rather than isolated code snippets, providing a more realistic software engineering context. Additionally, these games differ in how they handle working with actual code or the lack thereof. Some games approach this by presenting actual code snippets and asking quiz questions about them \cite{agrahari2020refactor4green,dos2019cleangame,baars2019codearena,elezi2016game}, while others abstract this away \cite{haendler2019card} or ask about code smells or refactoring techniques explicitly \cite{aljedaani2024boring}. While \textsc{CleanGame}  and \textsc{CodeArena}  allowed players to directly upload Java source and/or generate quiz questions, the game enables interaction with source code, though it serves as a feature rather than a core part of the game. This led us to select \toolname \cite{alomar2026refactoria}, as working directly with code is an integral part of the game. Each level features a code editor and has the player \textit{editing} and \textit{running} code that impacts the game. This is only possible with two unique features of \toolname. First, \toolname has its own custom language with Java-like syntax. Second, \toolname has a custom execution environment that interacts with the in-game environment. When players attempt to \textit{run} their code, it is sent through a tokenizer, parser, and backend interpreter, all created to understand the proposed custom language. The output is then passed to a frontend interpreter, which handles updating the game environment appropriately. These two components combined provide another benefit: we do not have to take the player out of the game and break their immersion. Whether that would mean that the player would need to open an IDE to write the code or open a dialogue window for them to upload their code. We also do not need to be concerned about the overhead of having a complete compiler packaged with \toolname.

%\vspace{-.3cm}
\section{\toolname Overview}
\label{Section:workflow}

%\subsection{Gameplay Loop Overview}

\begin{figure}[H]
  \centering
  \includegraphics[width=\columnwidth]{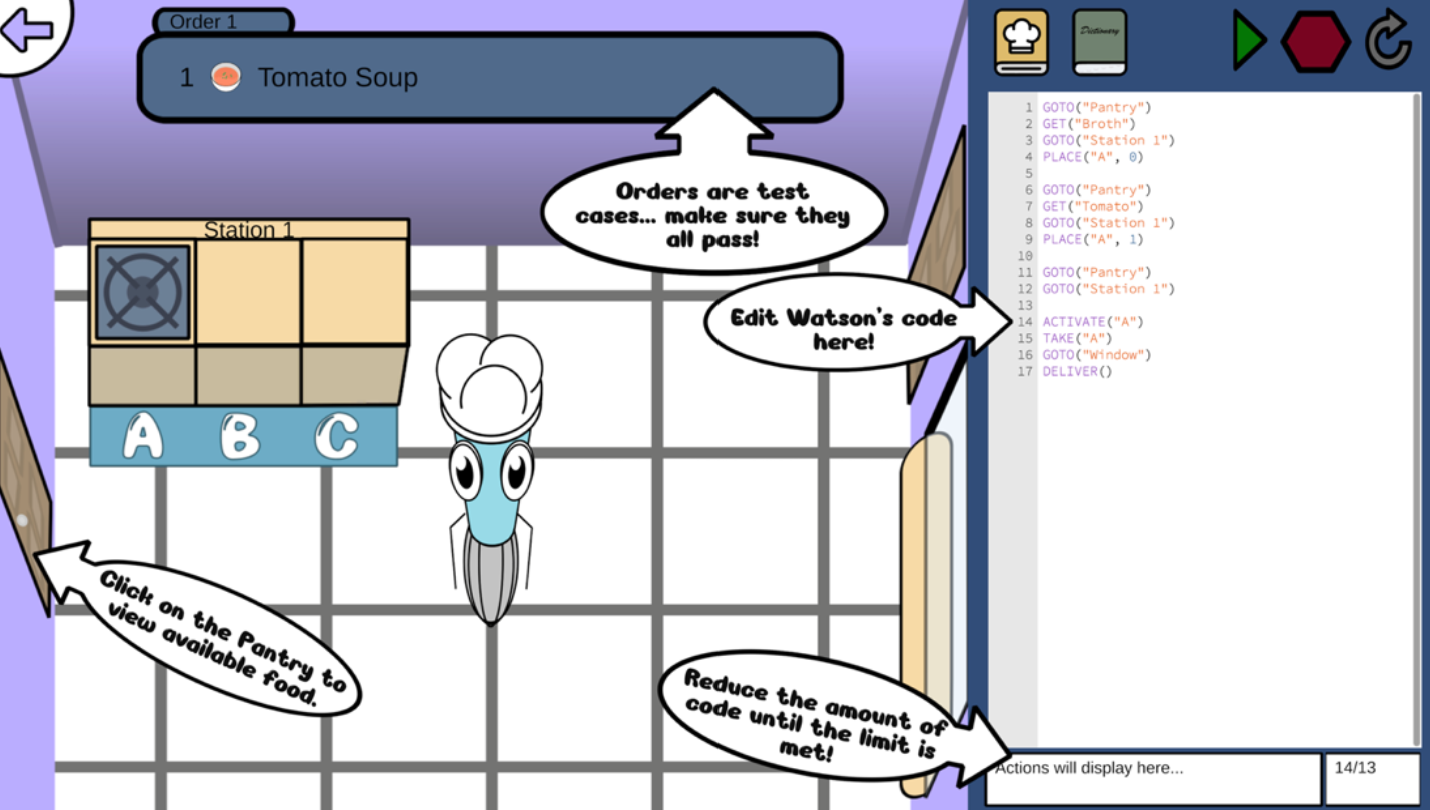}
  \caption{``Welcome to \textsc{Refactoria}'' puzzle.}
  \label{fig:puzzle9}
  %\vspace{-0.3cm}
 \end{figure}

The narrative of \toolname \cite{alomar2026refactoria} was designed to present players with an engaging story while also presenting them with a situation similar to something they might encounter in the real world. Keeping the story engaging encourages players to keep playing to hear the story. In addition, it is important that the storyline clearly sets up a natural need for refactoring. This was achieved by making Watson’s code base a legacy from the previous restaurant’s owner. This way, it is clearly shown in the narrative that the player is coming across this code for the first time, and that it needs to be updated and improved. This is like many situations in the industry, where programmers are given the task of updating a piece of software not originally written by them. Finally, each piece of narrative contains a small hint about how the player can solve the level’s puzzle (\eg in the level about \say{Speculative Generality}, %(when programmers add functionality in their code in case it is needed later), 
 the narrative states \say{\textit{The old owner frequently mentioned that he had intended to put new items on the menu, including squash soup (yuck!). That explains why he decided to leave it in the recipe book, but that recipe isn’t going to help us now}}). 
  Figure \ref{fig:puzzle9} shows an example of the first level.

\section{Experimental Setup and Execution}
\label{Section:Assignment}

\subsection{Goal and Research Question}
The purpose of this study is to validate \toolname to evaluate the acceptance and usefulness of the tool. The perspective comes from students who are willing to evaluate the feasibility of using the tool as a constructive aid and a continuous learning method during their programming activities. Based on our goal, we define the following research question (RQ): \textit{What is the perceived usefulness of \toolname?}

\subsection{Course Overview}

Object-oriented programming is an undergraduate course comprising two weekly lectures, each lasting 1 hour and 50 minutes. The course explores the foundations of object-oriented programming. Students were also given several hands-on assignments on data structures, testing, and related topics. The course deliverables consisted of 6 individual homework assignments, 14 quizzes, and 13 labs.

\subsection{Teaching Context and Participants}

The study involves one assignment in the object-oriented programming course. The course was taught at Stevens Institute of Technology. Before conducting the assignment, students have already learned about several code and design quality concerns: (1) code smells (teaching bad programming practices that violate design principles), and (2) code refactoring (teaching refactoring recipes that help improve software quality). This assignment introduces students to the fundamental concepts of code refactoring through an interactive, game-based learning environment. By engaging with the refactoring software, students will practice identifying and improving poor coding structures while maintaining program functionality.  It constituted 5\% of the final grade. It was due 14 days after the corresponding sessions. In total, 30 students completed the assignment. The respondents' programming experience ranged from 1 to 3 years, and their Java experience ranged from 1 to 2 years.

\subsection{Study Execution}

The assignment ensured that certain features were intuitive and the puzzles were doable. The format was designed for students to work through the game's levels and complete a survey. In summary, the students followed these steps: (1) install and launch \toolname, (2) play through the levels of the game. The game consists of multiple levels, each presenting a piece of poorly structured code, (3) identify code smells such as long methods, and duplicate code, (4) check the given constraints (\eg reducing the number of lines, improving readability, or eliminating redundancies), (5) use refactoring techniques, and ensure the refactored code maintains the original behavior and functionality, and (6) fill out the form.

\subsection{Survey Design} We follow the
guidelines proposed by Kitchenham and Pfleeger when designing our survey \cite{kitchenham2008personal}. The survey consisted of 36 questions that were divided into two parts. The first part of the survey includes demographic questions about participants. In the second part,
we asked about the (1) puzzle attempts, perceived ease, and difficulties (if any), (2)
  understandability of refactoring after completing the game, (3) challenges faced
when playing the game, (4) additional refactoring puzzles players would like to see, and (5) the application of refactoring techniques in future software development projects. We constructed
the survey to use open-ended questions, and multiple choice questions
with an optional \say{Other} category, allowing the respondents
to share thoughts not mentioned in the list. To increase the precision of our survey, we followed Smith's \etal guidelines \cite{smith2013improving}, and we targeted participants who have previously been exposed to coding. This process resulted in targeting 30  subjects
who are currently active developers. In
total, 24 participants provided their informed consent before participating in the study. %23 of the participants indicated that they are familiar with the general concept of refactoring (see Figure \ref{concept}). The coding experience of the respondents ranged from 1 to 3 years.

%\input{Tables/SurveyDesign}

%\vspace{-0.53cm}
\subsection{Data Analysis} Two authors independently reviewed the survey results. One author performed the data extraction and analysis independently, while the other reviewed and analyzed the extracted data. At the end, the authors met and discussed the survey results to reach a consensus.

\section{Results}
\label{Section:Result}

\begin{figure}[htbp]
        \centering
        \includegraphics[width=\linewidth, trim=8 75 20 20, clip]{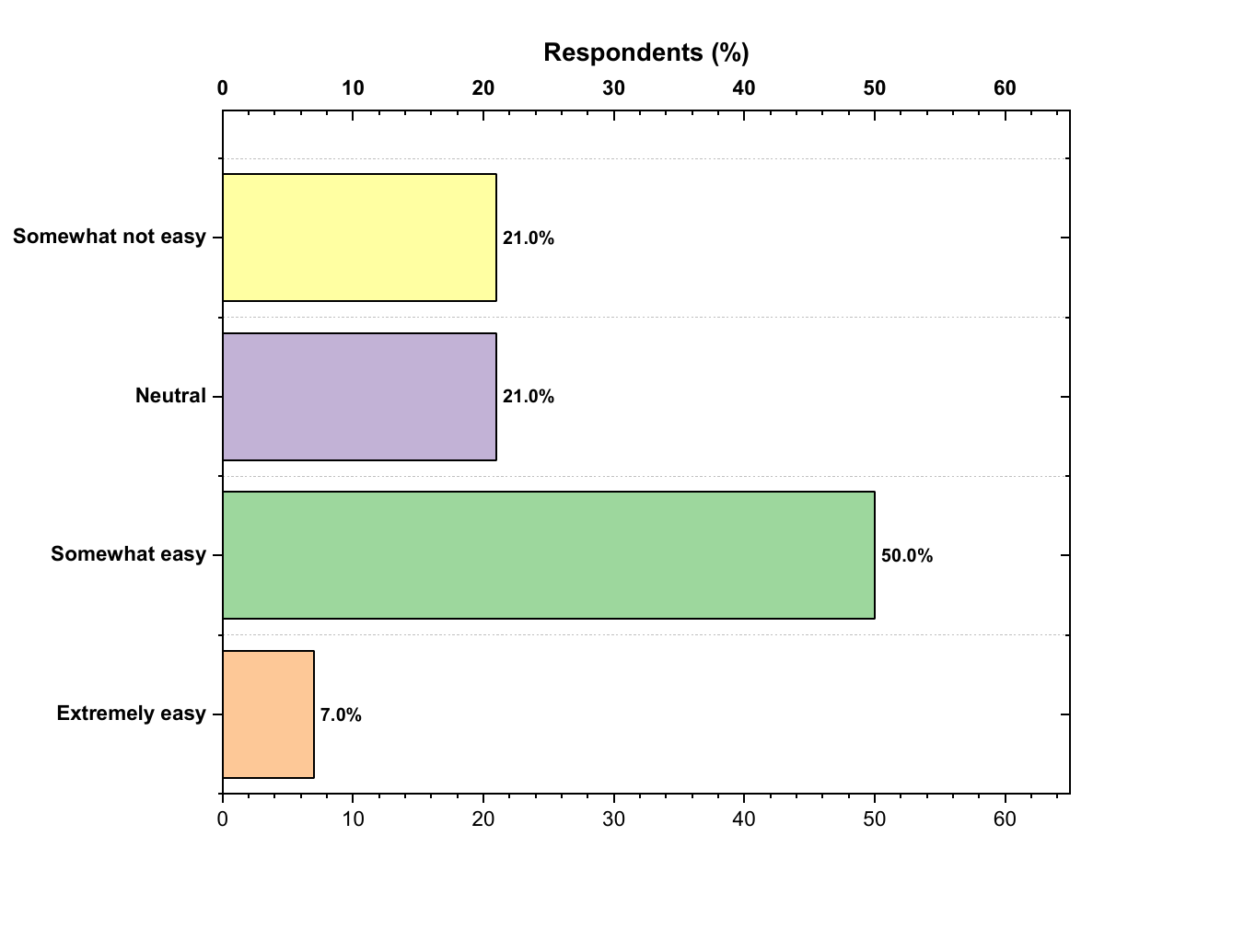}
    \caption{How easy was it to understand \toolname interface? (N=24)}
    \label{fig:easy}
    \vspace{-0.3cm}
\end{figure}

\begin{figure}[htbp]
        \centering
        \includegraphics[width=\linewidth, trim=4 75 20 20, clip]{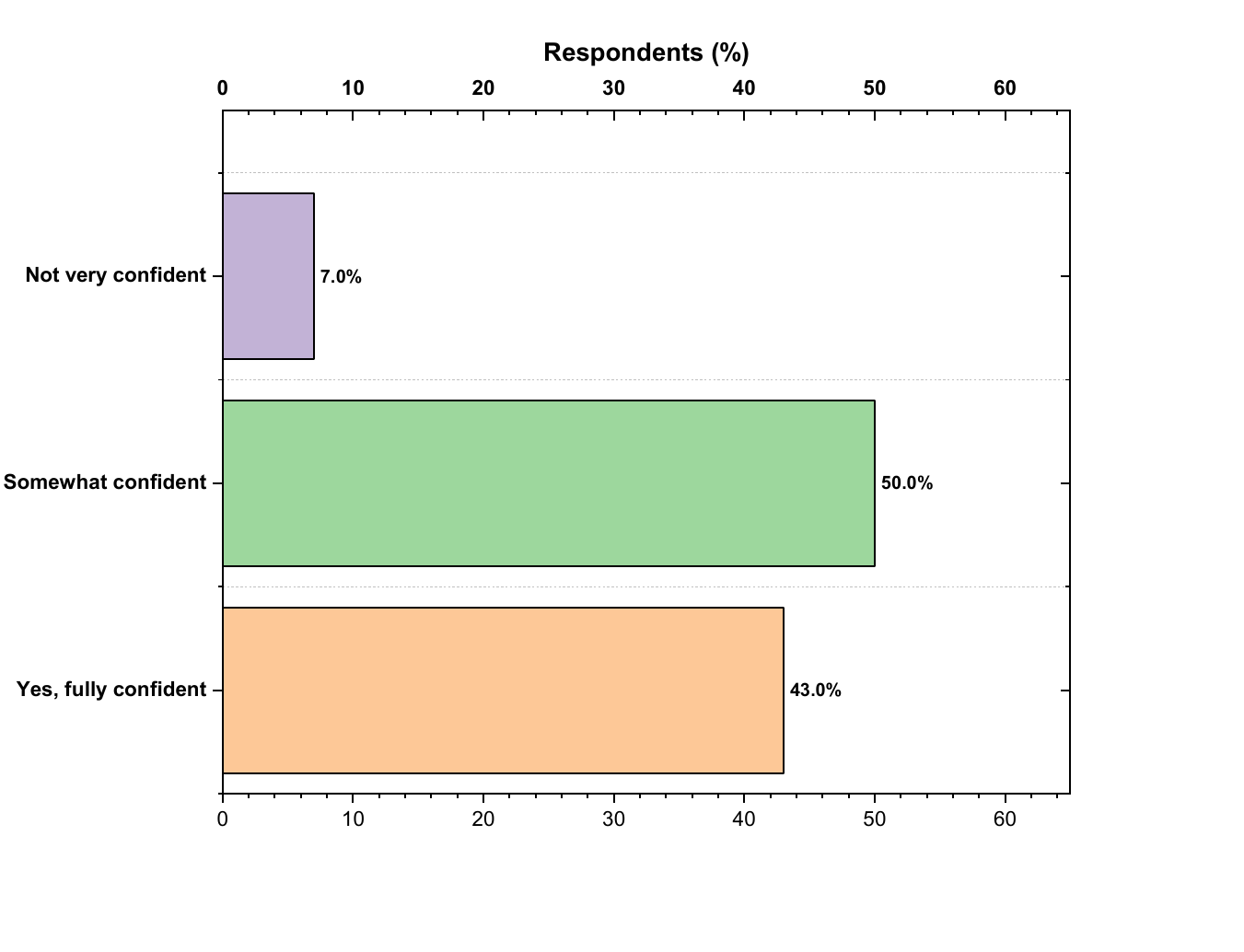}
    \caption{Did you feel confident performing refactoring on the source code after playing the game? (N=24)}
    \label{fig:confidance}
    \vspace{-0.3cm}
\end{figure}
\begin{figure*}[htbp]
    \centering
    % Row of 3 subfigures
    \begin{subfigure}[b]{0.32\textwidth}
        \centering
        \includegraphics[width=\linewidth, trim=20 50 20 40, clip]{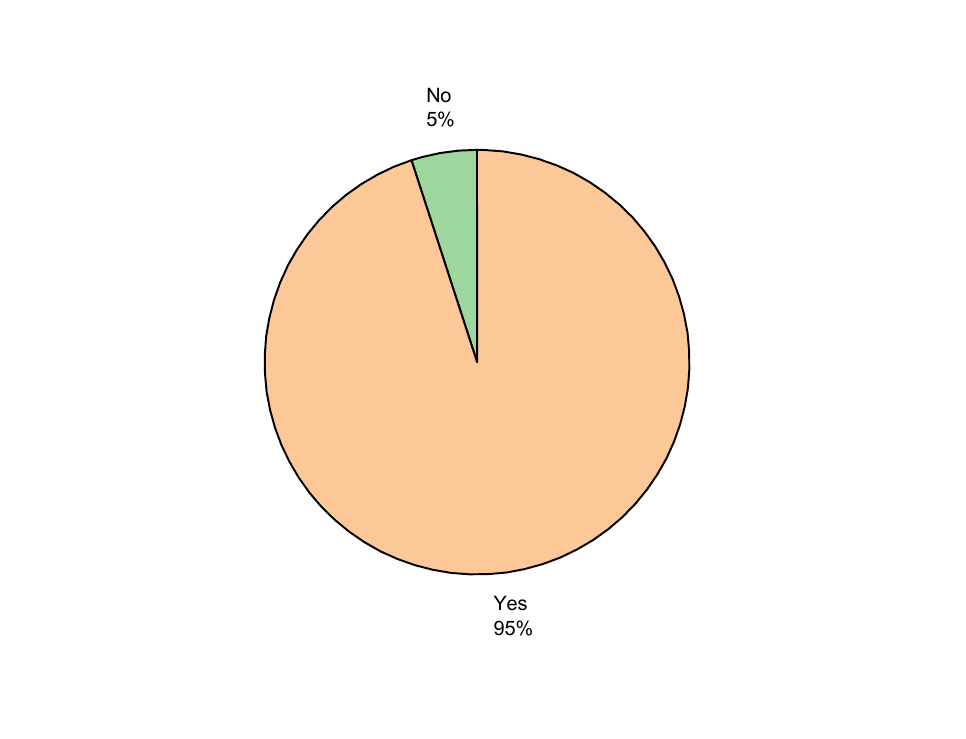}
        \caption{Are you familiar with the general definition of refactoring?}
        \label{concept}
    \end{subfigure}\hfill
    \begin{subfigure}[b]{0.32\textwidth}
        \centering
        \includegraphics[width=\linewidth, trim=20 50 20 40, clip]{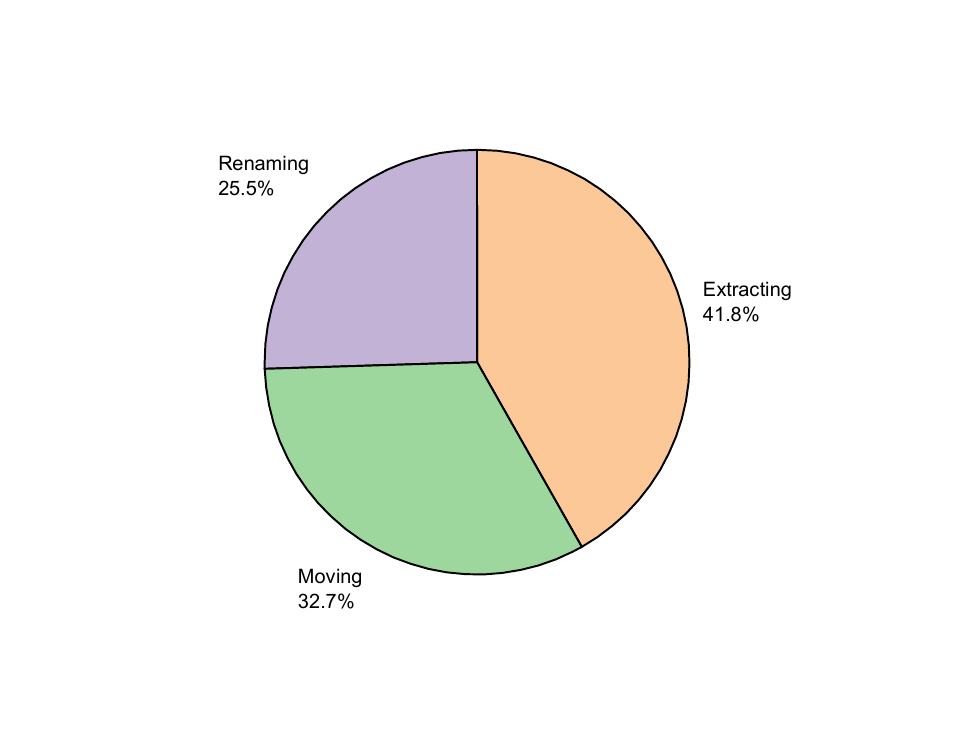}
        \caption{What key refactoring techniques do you feel most confident in applying?}
        \label{type}
    \end{subfigure}\hfill
    \begin{subfigure}[b]{0.32\textwidth}
        \centering
        \includegraphics[width=\linewidth, trim=20 50 20 40, clip]{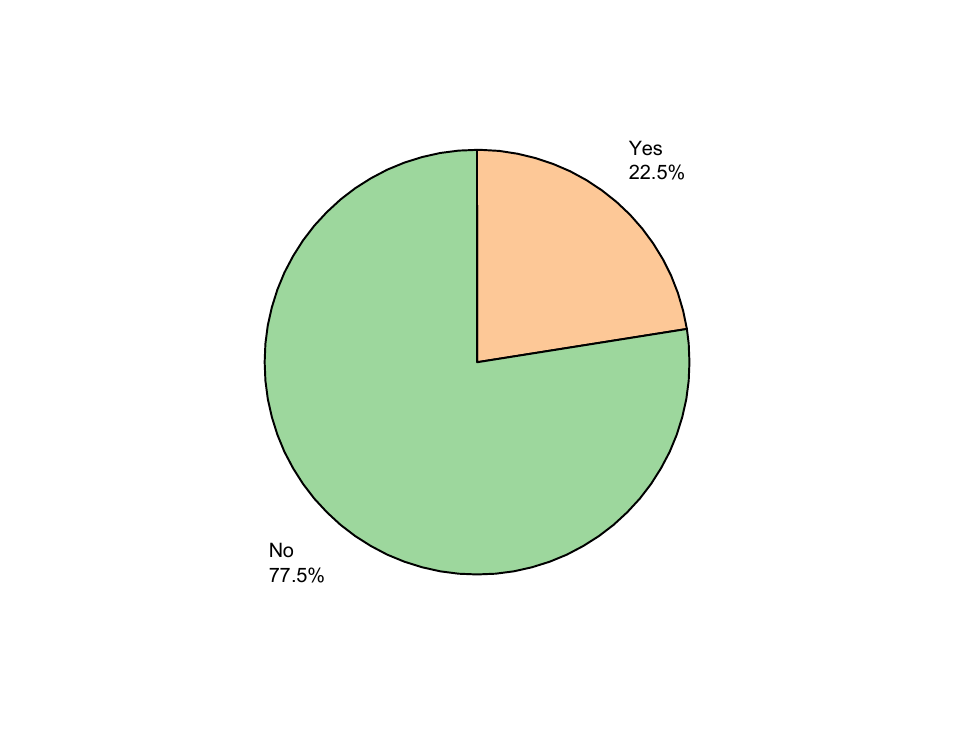}
        \caption{Were there challenges in understanding why refactoring was necessary?}
        \label{challenge}
    \end{subfigure}
    
    \caption{Participants' responses about refactoring: (a) definition, (b) techniques, and (c) challenges (N=24).}
    \label{fig:survey}
\end{figure*}

To establish a baseline for our evaluation, we first evaluated participants' knowledge of refactoring concepts. As shown in Figure \ref{concept}, 95\% of the participants were familiar with the concept, while only 5\% were not. %This baseline shows that the user study mainly assesses \toolname's effectiveness as an active practice for students who are already familiar with refactoring in theory, rather than its usefulness as an introductory tool for complete beginners.  
 Consequently, the subsequent findings regarding confidence and skill acquisition reflect the game's ability to bridge the gap between theoretical knowledge and practical application.

A primary factor in the success of an educational game is an intuitive interface that does not impede learning. Figure \ref{fig:easy} depicts the distribution of responses regarding the ease of understanding \toolname's interface. The results lean positive, with 67\% of respondents rating the interface as either ``Extremely easy''  or ``Somewhat easy''.  However, a minority found it ``Somewhat not easy'', and few participants remained ``Neutral.'' This indicates that although the game interface is accessible to most users, understanding the game's mechanics involves a learning curve.

The primary pedagogical goal of \toolname is to empower students to refactor source code with confidence. Figure \ref{fig:confidance} shows a strong positive outcome in this regard. Among the respondents, the majority reported feeling confident after playing the game, with 50\% of participants choosing ``Yes, fully confident'' and 43\% choosing ``Somewhat confident.'' 7\% of participants reported feeling ``Not very confident.''

\begin{figure}[htbp]
        \centering
        \includegraphics[width=\linewidth, trim=20 75 20 20, clip]{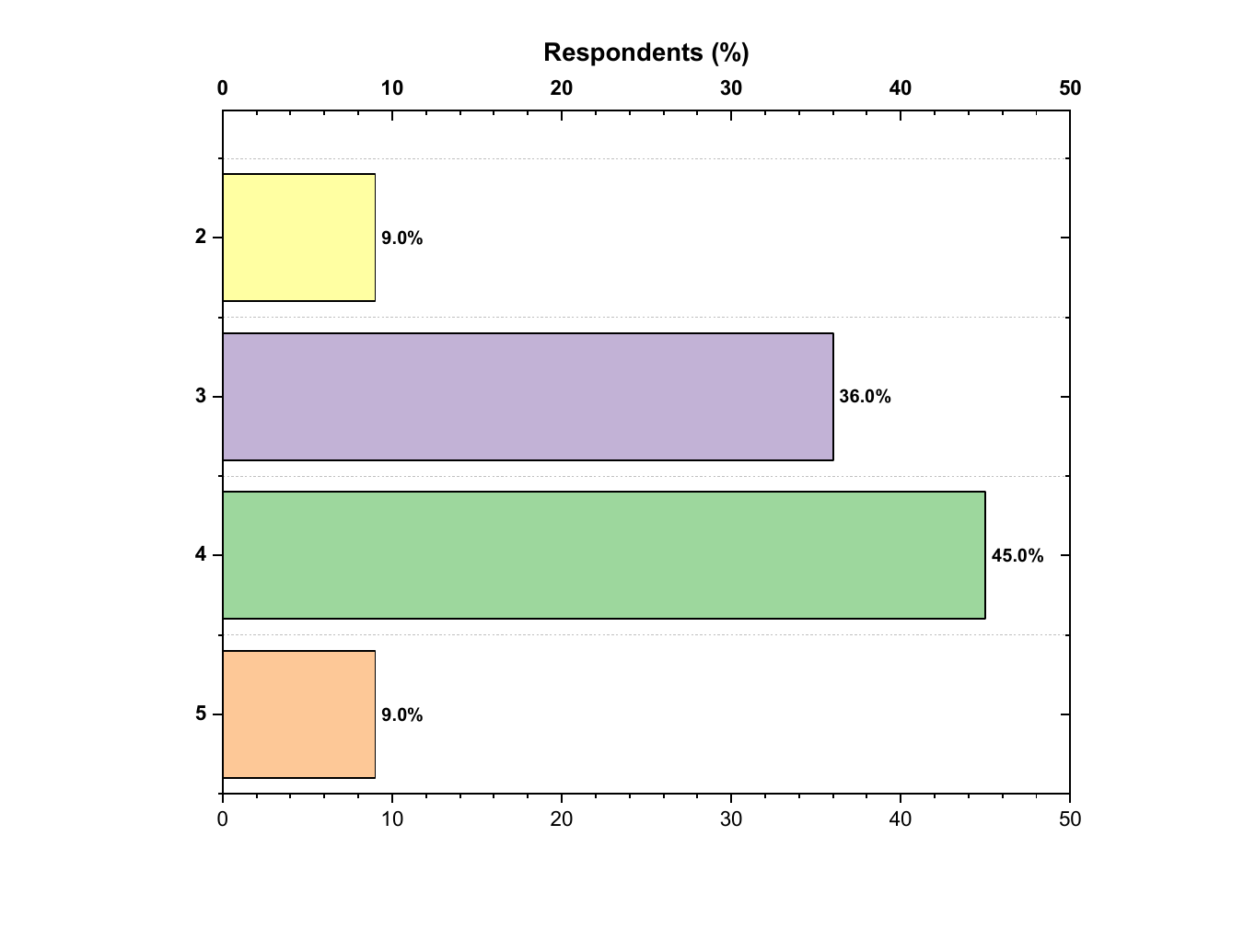}
    \caption{To what extent did the Java-like syntax refactoring logic in Refactoria mirror the decision-making process you use when refactoring in Java? (N=24)}
    \label{fig:scale}
    \vspace{-0.3cm}
\end{figure}

For an educational game to be effective, its mechanics must be transferred to real-world coding tasks. Figure \ref{fig:scale} shows how well participants felt the Java-like syntax and logic in \toolname mirrored their actual decision-making processes when refactoring in Java, rated on a 1–5 scale. The responses indicate moderate to strong authenticity, with a mean score of 3.55 and a median of 4.0. More than half of the respondents rated the mirroring positively (scores of 4 or 5). While these results confirm that the game effectively captures the essence of Java refactoring, the presence of neutral and lower scores suggests that some abstractions required for gameplay may differ from IDE-based refactoring workflows.

% Furthermore, we evaluated whether participants understood the underlying motivation for refactoring. Figure \hl{\#} shows that 76\% of participants reported not experiencing challenges understanding why refactoring was necessary during gameplay. This is an encouraging result, indicating that \toolname effectively communicates code smells and the value of structural improvements. However, the 24\% who did experience challenges highlight an area for pedagogical improvement. Although the game effectively teaches refactoring, future versions can include immediate feedback that clearly connects mechanical actions to improvements in code quality.

\begin{figure}[htbp]
        \centering
        \includegraphics[width=\linewidth, trim=10 40 0 0, clip]{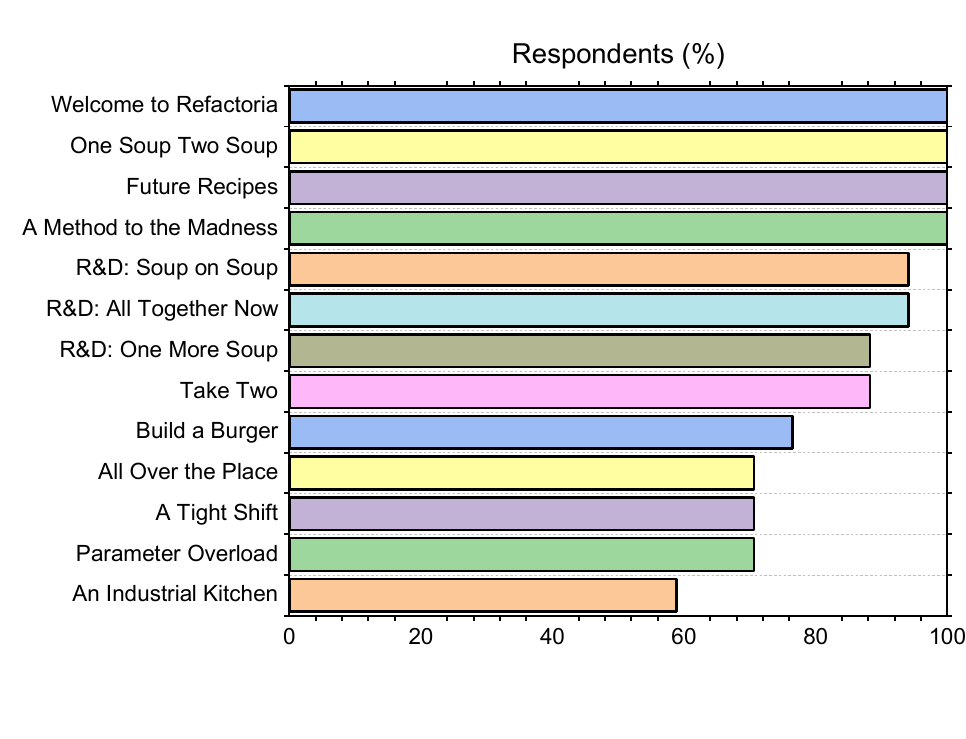}
    \caption{Distribution of participants who successfully solved each puzzle (N=24).}
    \label{fig:puzzle}
    \vspace{-0.3cm}
\end{figure}

The survey responses show that 83\% of participants responded positively to the open-ended question \say{How has your understanding of refactoring changed after completing the game?}. 6 participants answered that the game allowed them to practice the refactoring concepts, and 4  participants mentioned learning new refactoring concepts. A few participants shared their opinion: P16 states that, \say{\textit{I'm not as scared to go through my old code and refactor because the game and its silliness really helped me get over myself.}}, and P7 mentions that, \say{\textit{After completing the game, I understand refactoring more as improving the structure, readability, and reusability of code without changing what the program actually does}}. 2 participants answered that they already had experience with refactoring, and so this game did not teach them anything new. However, this game is intended as an introduction to refactoring, so we anticipate that anyone who has prior knowledge and experience would most likely not play the game. There were no examples of someone without prior experience with refactoring answering that they had not learned anything. Therefore, we can conclude that the participants viewed the game as effective in teaching refactoring.  In response to the question \say{How do you see yourself applying refactoring techniques?}, 7 participants mentioned continuous refactoring or refactoring before submitting to the repository. Furthermore, 12 participants (50\%) mentioned making their code more efficient or less repetitive. Some participants share their opinion during the survey: P1 states that, \say{\textit{This was helpful for identifying what parts of the code could be refactored to cut down on the number of lines, which will be helpful for writing my future programming assignments.}}, P2 states that, \say{\textit{I would probably look over my finished code one more time to figure out if I could refactor the code a bit more because in some of the puzzles it isn't obvious at first where I could simplify the code even more.}}, and P3 states that, \say{\textit{i definity think that i would be applying it to make code cleaner and more readable}}. Because efficiency and repetition were a major part of the first few puzzles, we can conclude that the game effectively teaches players these aspects of refactoring. We inquired about participants' views on key refactoring techniques they feel most confident in applying. Figure \ref{type} illustrates that \say{Extracting} was mentioned by 41.8\% participants, \say{Moving} by 32.7\%, and \say{Renaming} by 25.5\%. When we asked the participants, \say{Were there any challenges in understanding why refactoring was necessary?}, 77.5\% of participants indicated that there are no challenges in understanding the importance of refactoring after playing the game (see Figure \ref{challenge}). \textcolor{black}{For participants indicating a challenge, few stated the reason. For example, P12 notes, \say{\textit{Yes, there were challenges in understanding why refactoring was necessary at first. Initially, the code “worked,” so it wasn’t obvious why changes were needed. However, as constraints like module limits and efficiency were introduced, it became clear that poorly structured code leads to unnecessary complexity and inefficiency. Refactoring helped simplify the logic, reduce repetition, and make the code easier to manage, which highlighted its importance.}}, and P7 notes, \say{\textit{At first, some of the code already worked, so it was not always obvious why changing it was necessary. I gradually understood that refactoring was important for reducing duplication, improving readability, and making the code easier to maintain.}}} \textcolor{black}{Figure \ref{fig:puzzle} informs the difficulty of each level in \toolname, relative to each other. There are four levels that are at 100\% completion rate:  ``Welcome to \toolname'', ``One Soup Two Soup'', ``Future Recipes'', and ``Method to the Madness''. High completion rates were also observed for ``R\&D: Soup on Soup'' and ``R\&D: All Together Now'' (94\%), followed by ``R\&D: One More Soup'' and ``Take Two'' (88\%). ``Build a Burger'' was correctly identified by 76\% of respondents, while ``All Over the Place'', ``A Tight Shift'', and ``Parameter Overload'' each achieved 71\%. The lowest completion rate was recorded for ``An Industrial Kitchen'' (59\%). We see a steady decrease in completion rate with no difficulty spikes. }
%there are four levels that are at or below a 50% completion rate. The first three are clumped together: “R&D: All Together Now”, “R&D: One More Soup”, and “Take Two”. For these levels there is a completion rate of [placeholder]%, [placeholder]%, and [placeholder]% respectively. From the level before this cluster, “R&D: Soup on Soup”, there is a [placeholder]-percentage-point decrease in completion rate. The level after the cluster, “Take Two”, there is a [placeholder]-completion-point increase. The last level in this group is “A Tight Shift” with a completion rate of [placeholder]%. Compared to its neighboring levels, there is a [placeholder]-percentage-point decrease from “All Over the Place” and a [placeholder]-percentage-point increase to “Parameter Overload”. }.

%illustrates the number of participants categorized according to their ability to solve the puzzle. As can be seen, the majority of the participants were able to solve 8 puzzles. %It is important to note that \say{Welcome to \toolname} was introduced later as a response to a playtest comment, so only 3 participants tried it out and were able to solve it.
  
  Moreover, when asked, \say{Describe your experience interacting with the in-game custom language. Was it intuitive?}, most of the participants (14 respondents (58\%)) indicated that it was intuitive. Others revealed that it would be better using an actual language that they can use their prior knowledge on, and after seeing the code execute in the first puzzle, it becomes easy to pick up on the syntax of the language. The main development that came as a direct result of this playtesting was %the addition of \say{Welcome to \toolname}, an easy puzzle to start the game and get the player adjusted to the concepts of modules and stations. Another important development was a
   the addition of tips and tricks to the story blurbs for \say{Welcome to \textsc{Refactoria}} to help the player out in the early stages of the game and reduce confusion.  Further, when asked about additional refactoring puzzles they would like to see, participants mentioned specific refactoring operations such  \say{Inline} and \say{Rename} refactoring, and others indicated recursive methods. Three participants stated, P9: \say{\textit{I would like to see more extracting challenges with different complexities/types of code!}}, P13, \say{\textit{Challenges involving recursive methods could be cool.}}, and P15, \say{\textit{Maybe inline method challenges if possible}}. Another participant stated, P7, \say{\textit{I can't think of any additional refactoring challenges that were not already implemented in this game}}. In addition, participants were asked whether the architectural principles learned in the game are directly applicable to object-oriented languages like Java.  The applicability of principles to object-oriented languages was confirmed in 85\% of the responses. P6 notes that, \say{\textit{Yes, I think these principles are directly applicable to object-oriented languages like Java because Java programs also benefit from reducing duplication, organizing responsibilities more clearly, and making code easier to read, test, maintain, and extend.}},  P8 states that, \say{\textit{I do think so, as almost all the problems related to some `object' in this case the food materials and involved fixing the constructors or methods related to the food materials which is broadly similar to how coding for java objects}}, and P9 mentions that, \say{\textit{Yeah Yes, because the game's logic directly mirrors core principles like abstraction and encapsulation. By building reusable functions, I’m implementing the Don't Repeat Yourself principle to keep code maintainable. Furthermore, spreading tasks across stations follows the Single Responsibility Principle, ensuring each "module" has one specific job and preventing the creation of unmanageable objects}}.

% \begin{figure}[htbp]
%         \centering
%         \includegraphics[width=\linewidth, trim=8 75 20 20, clip]{Images/sigcse-bar-confidence.pdf}
%     \caption{Did you feel confident performing refactoring on the source code after playing the game?}
%     \label{fig:confidance}
%     \vspace{-0.3cm}
% \end{figure}

% 20 → trim from left
% 95 → trim from bottom
% 20 → trim from right
% 40 → trim from top

%\vspace{-.6cm}
\section{Reflection}
\label{Section:Reflection}

\noindent{\textbf{What instructors can do.}} Our work provides the following actionable guidance for designing refactoring games. (1) Use constraints such as test preservation to encourage students to reason about refactoring trade-offs. (2) Provide automated feedback, such as test execution, to assist players in understanding the impact of refactoring action. (3) Integrate the game as a complement (not a replacement) to traditional teaching instructions. \textcolor{black}{In addition, instructors can assess improvements in the quality of students' programming submissions by comparing code quality metrics before and after the use of \toolname, thereby evaluating its impact on learning outcomes.}

\noindent{\textbf{What went right.}}   \toolname encourages students to actively identify, analyze, and refactor code smells through trial-and-error in an engaging environment. \toolname progressively introduces code smells, starting with simple cases and progressing to complex scenarios. We observe that students reported feeling a sense of accomplishment when solving \toolname game puzzles, attributing their success to their own skills. Additionally, the test cases help students identify effective refactorings and reinforce the importance of preserving behavior. 

\textcolor{black}{In every level of \toolname, a functional code snippet is given to them which they must read and understand before solving the puzzle. This experience is directly applicable to the real world because developers are often faced with the challenges of working on existing code using packages or technologies that they are not already familiar with.} 

\textcolor{black}{Often, the initial code given in the levels was deliberately designed to be hard to read. The solutions are required to be efficient, and as a byproduct, they are often much easier to read. This game gives players first-hand experience dealing with hard-to-read code and the easy-to-read equivalent. Therefore, the student developers in the future should have an easier time recognizing what hard-to-read code looks like and how to fix it. }

\textcolor{black}{Using a custom language was helpful. We can confirm that the learning from this game was agnostic of the student developers' previous experience using any one specific language.}

%\toolname constraints prompted students to focus on structural improvements in the code.
\noindent{\textbf{What went wrong.}} A few players focus on game constraints rather than code maintainability, emphasizing a misalignment between the game mechanics and learning objectives. \textcolor{black}{Additionally, a few students who are familiar with the concept of refactoring indicate that they do not learn new techniques, but the game reinforces the concept. P11 states, ``\textit{I was pretty familiar with it before, but this just gave me more practice}’’.  More advanced levels can be implemented to keep experienced students engaged and challenged.}  %\textcolor{black}{Further, the series of three levels labelled R\&D is intended to be a relate set of three puzzles. The major dropoff from ``R\&D: Soup on Soup'' to ``R\&D: All Together Now'' in Figure \ref{fig:puzzle} shows that the lesson from ``Soup on Soup'' was not effective because students were generally not easily able to transfer that learning to ``All Together Now.''}

\noindent{\textbf{What could be improved in future iterations.}} Future versions of \toolname should incorporate reflection prompts to allow students to justify refactoring decisions beyond completing the level. This is essential as explaining \textit{why} refactoring improved/degraded quality is as important as understanding \textit{what} refactoring applied, as both strengthen the learning outcomes. \textcolor{black}{Additionally, if a player of \toolname makes a genuine attempt to solve the puzzle, and they cannot figure it out, there is no venue for help. Ideally, a future iteration of the game would have a hint system. Perhaps, the game would time how long a player has spent on a level, and if enough time has elapsed, a hint is offered. Further, one critical area that was not considered during the development of \toolname is accessibility. According to the World Health Organization, as of 2023, 1 in 6 people has some disability\footnote{\url{https://www.who.int/news-room/fact-sheets/detail/disability-and-health}}. It is not a stretch that individuals with disabilities would play \toolname, especially in an academic environment. There are relatively simple changes and features that can be added. The built-in code editor features syntax highlighting but is set to a single theme that cannot be changed. Introducing additional options or allowing players to make their own theme, would help cater to players who are colorblind or need high-contrast colors to write code. An option to use dyslexia-friendly fonts, like OpenDyslexic\footnote{\url{https://opendyslexic.org/}}, would help dyslexic players interpret and focus on text elements throughout the game and while coding. More complex changes would be supporting and making optimizations for screen readers, for blind or low-vision players, and voice control, for players with motor impairments. For both assistive technologies, this task would be much more intensive as \toolname would need to be made interoperable with external software it was never designed to interact with. User testing could be pursued alongside these efforts to determine how effective these changes are and other areas where accessibility could be improved.}

\section{Conclusion}
\label{Section:Conclusion}

This work presents \toolname, an interactive gamified tool that teaches the basic concepts of code smells and refactoring. It is designed to give semi-experienced programmers an introduction to refactoring. The players play as an expert chef with their sidekick Watson the Whiskbot to refactor Watson's instructions into efficient, readable, and easily maintainable code. While making fun recipes, players learn better coding etiquette. In each of the 13 levels, players are presented with a code-based puzzle that, although it passes the test cases, contains a variety of code smells. Then, they are tasked with applying refactoring techniques to ensure that the code meets one or two additional requirements designed to teach better coding practices. We conduct a user study to assess the game's usefulness. Based on this analysis, we conclude that participants found the game helpful in learning refactoring.
%rated \toolname as effective for teaching refactoring. 
  %Future work involves extending the game to support additional code smells and their associated refactoring techniques.

%AlOmar, Eman Abdullah and Van Hise, David and Pachucki, Stephen and Lundquist, Nerissa and Engelbart, Christopher

\bibliographystyle{ACM-Reference-Format}
%\bibliographystyle{abbrv}
%\bibliography{IEEEabrv,sample-base}
\bibliography{sample-base}

\end{document}